\documentclass[12pt,letterpaper]{article}
\usepackage{amsmath,amssymb,calc}
\usepackage{color}
\usepackage{graphicx}
\usepackage[nosort]{cite}

\makeatletter
\renewcommand\section{\@startsection {section}{1}{\z@}%
                                   {-3.5ex \@plus -1ex \@minus -.2ex}%nn
                                   {2.3ex \@plus.2ex}%
                                   {\normalfont\large\bfseries}}
\renewcommand\subsection{\@startsection{subsection}{2}{\z@}%
                                     {-3.25ex\@plus -1ex \@minus -.2ex}%
                                     {1.5ex \@plus .2ex}%
                                     {\normalfont\bfseries}}
\makeatother

\let\non\nonumber

\newcommand{\bea}{\begin{eqnarray}}
\newcommand{\eea}{\end{eqnarray}}
\newcommand{\be}{\begin{equation}}
\newcommand{\ee}{\end{equation}}
\newcommand{\bma}{\begin{pmatrix}}
\newcommand{\ema}{\end{pmatrix}}

\newcommand{\bF}{{\mathbb F}}

\newcommand{\cA}{{\cal A}}

\newcommand{\cF}{{\cal F}}
\newcommand{\cG}{{\cal G}}
\newcommand{\cI}{{\cal I}}

\newcommand{\cM}{{\cal M}}
\newcommand{\cP}{{\cal P}}
\newcommand{\cR}{{\cal R}}
\newcommand{\cS}{{\cal S}}
\newcommand{\cU}{{\cal U}}
\newcommand{\cV}{{\cal V}}
\newcommand{\cX}{{\cal X}}

\newcommand{\db}{{\mathsf b}}
\newcommand{\dc}{{\mathsf c}}
\newcommand{\dX}{{\mathsf X}}
\newcommand{\bhs}{{\db^*}}
\newcommand{\chs}{{\dc^*}}

\newcommand{\rf}{{\rm f}}
\newcommand{\rs}{{\rm s}}
\newcommand{\fa}{{f_a}}
\newcommand {\bflam}{{\boldsymbol\lambda}}
\newcommand{\gb}{{\gamma_b}}
\newcommand{\gc}{{\gamma_c}}

\newcommand{\gp}{{\gamma_\psi}}

\newcommand{\lam}{\lambda}

\newcommand{\da}{\dot{a}}

\newcommand{\pref}[1]{$(\ref{#1})$}

\def\eg{{\it e.g.}}
\def\ie{{\it i.e.}}

\def\IZ{\relax\ifmmode\mathchoice
{\hbox{\cmss Z\kern-.4em Z}}{\hbox{\cmss Z\kern-.4em Z}}
{\lower.9pt\hbox{\cmsss Z\kern-.4em Z}} {\lower1.2pt\hbox{\cmsss
Z\kern-.4em Z}}\else{\cmss Z\kern-.4em Z}\fi}
\def\IR{\relax{\rm I\kern-.18em R}}
\def\One{{\hbox{ 1\kern-.8mm l}}}

\def\tr{{\rm tr\,}}
\def\Tr{{\rm Tr\,}}

\newlength{\bredde}
\def\slash#1{\settowidth{\bredde}{$#1$}\ifmmode\,\raisebox{.15ex}{/}
\hspace*{-\bredde} #1\else$\,\raisebox{.15ex}{/}\hspace*{-\bredde}
#1$\fi}

\newsavebox{\zzzbar}
\sbox{\zzzbar}
  {\setlength{\unitlength}{0.9em}
  \begin{picture}(0.6,0.7)
  \thinlines
  \put(0,0){\line(1,0){0.6}}
  \put(0,0.75){\line(1,0){0.575}}
  \multiput(0,0)(0.0125,0.025){30}{\rule{0.3pt}{0.3pt}}
  \multiput(0.2,0)(0.0125,0.025){30}{\rule{0.3pt}{0.3pt}}
  \put(0,0.75){\line(0,-1){0.15}}
  \put(0.015,0.75){\line(0,-1){0.1}}
  \put(0.03,0.75){\line(0,-1){0.075}}
  \put(0.045,0.75){\line(0,-1){0.05}}
  \put(0.05,0.75){\line(0,-1){0.025}}
  \put(0.6,0){\line(0,1){0.15}}
  \put(0.585,0){\line(0,1){0.1}}
  \put(0.57,0){\line(0,1){0.075}}
  \put(0.555,0){\line(0,1){0.05}}
  \put(0.55,0){\line(0,1){0.025}}
  \end{picture}}

\newcommand{\bra}[1]{\langle{#1}|}
\newcommand{\ket}[1]{|{#1}\rangle}

\newcommand{\vev}[1]{\langle{#1}\rangle}
\newcommand{\ena}{\end{eqnarray}}
\newcommand{\beqa}{\begin{eqnarray}}
\newcommand{\eeqa}{\end{eqnarray}}

\newcommand{\half}{\frac{1}{2}}

\def\eff{{\rm eff}}
\def\gstr{g_{s}}
\def\lstr{\ell_{s}}
\def\lpl{\ell_{p}}
\def\mpl{m_p}

\def\tilsig{{\widetilde\Sigma}}
\def\tilomega{{\widetilde \omega}}

\newfont{\goth}{ygoth.tfm scaled 1200}                   % gothic font (usual)
\numberwithin{equation}{section}

\begin{document}
\begin{titlepage}

\begin{center}

%{September 26, 2005}
\today \hfill         \phantom{xxx} \hfill EFI-12-12

\vskip 2 cm {\Large \bf Chern-Simons EM-flation}
\vskip 1.25 cm {\bf  
Emil Martinec,$^{a,b,}$\footnote{ejmartin@uchicago.edu} 
Peter Adshead,$^{a,c,}$\footnote{adshead@kicp.uchicago.edu} 
{\rm and} 
Mark Wyman$^{a,c,d,}$\footnote{markwy@oddjob.uchicago.edu}
}
\non\\
{\vskip 0.5cm  
$^{a}${\it Enrico Fermi Institute,} $^{b}${\it Department of Physics}\non\\ \vskip 0.2cm
$^{c}${\it Kavli Institute for Cosmological Physics}\non\\ \vskip 0.2cm 
$^{d}${\it Department of Astronomy and Astrophysics}\non\\ \vskip 0.2cm
 {\it University of Chicago, Chicago, IL 60637, USA}\non\\ \vskip 0.2cm

}

\end{center}
\vskip 2 cm

\begin{abstract}
\baselineskip=18pt

We propose a new, generic mechanism of inflation mediated by a balance between potential forces and a Chern-Simons interaction.  Such quasi-topological interactions are ubiquitous in string theory.  In the minisuperspace approximation, their effect on the dynamics can be mapped onto the problem of a charged particle in an electromagnetic field together with an external potential; slow roll arises when the motion is dominated by the analogue of `magnetic drift'.  This mechanism is robust against radiative corrections.  We suggest a possible experimental signature which, if observed, might be considered strong evidence for string theory.

\end{abstract}

\end{titlepage}

%\tableofcontents

%%%%%%%%%%%%%%%%%%%%%%%%%%%%%%%%%%%%%%%%%%%%%%%
%%%%%%%%%%%%%%%%%%%%%%%%%%%%%%%%%%%%%%%%%%%%%%%

\section{Introduction}

Inflation%
~\cite{Guth:1980zm}
is at the same time a spectacular phenomenological success, and an enduring theoretical challenge.  It seems clear that the universe underwent a period of rapid expansion in its early history; however, it seems rather difficult to engineer a field-theoretic model which sustains a sufficiently extended inflationary phase.

Typical models of inflation involve the evolution of an effective scalar field, the inflaton, along its effective potential.  If the motion along the potential is sufficiently slow, so that the energy density is dominated by the inflaton potential energy over all other sources, the local value of the potential acts as an effective cosmological constant, slowly evolving in time.  The notion of `sufficiently slow' is quantified by the slow-roll parameters of the Hubble scale $H=\dot a/a$
\bea
\label{slowroll}
\epsilon_H &=& -\frac{\dot H}{H^2}  \nonumber\\
\eta_H &=& -\frac{\ddot H}{2H\dot H} \ .
\eea
The conditions $\epsilon_H,\eta_H\ll 1$ define the period of slow-roll inflation; one needs this era to last long enough so that the scale factor $a$ undergoes $60$ or more e-foldings of exponential expansion.  

There are a number of mechanisms that have been proposed to achieve such an extended inflationary era:
\begin{description}
\item{1.} Make the inflaton potential $\cV$ extraordinarily flat%
~\cite{Albrecht:1982wi,Linde:1981mu},
so that $H^2\sim\cV/3\mpl^2$ changes very slowly over a large region of field space.  Unfortunately, scalar field potentials are subject to strong radiative corrections, making such a flat potential generically unnatural.  The potential can be protected through a symmetry, for instance making the inflaton a pseudo-Goldstone boson as in `natural' inflation%
~\cite{Freese:1990rb}; however, this approach has its own set of issues -- typically one cannot arrange the periodicity of the axion to be large enough to sustain inflation long enough%
~\cite{Banks:2003sx}.%
\footnote{Models employing axion monodromy%
~\cite{McAllister:2008hb}
suggest a way around this issue.}
%Freese:2008if,
\item{2.} Engineer the kinetic term so that the inflaton motion has a limiting velocity and/or small sound speed%
~\cite{ArmendarizPicon:1999rj,Silverstein:2003hf}.  
Here again, the form of the kinetic energy needed is sensitive to quantum corrections.
%~\cite{Chen:2008hz}.
\item{3.} Arrange for some form of `friction' or damping to impede the evolution of the inflaton%
~\cite{Berera:1995ie},
%,Kofman:2004yc,Green:2009ds},
for instance via particle production.  These approaches again involve a degree of fine-tuning, since the produced particles mustn't dominate the energy density while being sufficiently produced to overcome the dilution resulting from the expansion, and at the same time achieving the needed damping; there can also be issues with scale dependence of perturbations.
\end{description}
In general, there is a consistency issue -- the `eta problem' -- as to why quantum corrections to the effective action of the inflaton $\phi$ don't contribute  $\Delta\eta=\Delta m_\phi^2/3H^2\sim O(1)$.  Supersymmetry, which can protect scalar masses, doesn't help here -- $H$ is an effective scale of supersymmetry breaking.

This article introduces a fourth general mechanism of inflation, which we call `EM-flation'.%
\footnote{We would have called it `Chern-Simons inflation' but that name appears to be taken%
~\cite{Alexander:2011hz}.}
This term refers to both its intrinsic reliance on the electromagnetic fluxes of antisymmetric tensor (AST) gauge fields, and to a strong analogy with the forced dynamics of a charged particle in a magnetic field.  The basic idea is that a Chern-Simons term, linear in time derivatives, can play the same role in field space dynamics that a magnetic field does for a point particle.  For the particle in a strong field, other force terms in the equations of motion balance against the magnetic force rather than kinetic terms.  Similarly, in the presence of a large Chern-Simons coupling, the resulting field motion is the direct analogue of the `magnetic drift' of a charged particle in a strong magnetic field.  One need assume no special features of either the kinetic energy or the potential -- the kinetic term plays only a minor role, and potentials can be steep; all that is required is to be able to make the strength of the Chern-Simons interaction sufficiently large, so that `magnetic drift' is slow, in order to generate a long inflationary era.  Because no special property of the effective action is assumed -- other than a large Chern-Simons coupling, whose coefficient is quantized -- the eta problem should be absent.

Recently, a model of `chromo-natural' inflation%
~\cite{Adshead:2012kp,Adshead:2012qe}
was proposed by two of the authors; as we shall see, this model is the first example of this new inflationary mechanism.
Let us summarize the chromo-natural inflationary model.  The starting point is natural inflation%
~\cite{Freese:1990rb}, 
where the inflaton is an axion field $\cX$.  The axion is assumed to have a potential generated by some unspecified non-perturbative effects, and interacts with a weakly-coupled $SU(2)$ gauge field via the usual topological density 
\be
\label{chiFF}
\lambda\left(\frac{\cX}{\fa} \right) \tr [\cF\wedge \cF] \  .  
\ee
Under the assumption of a texture in the gauge field background, conformal to the spatial dreibein,%
\footnote{More generally, the ansatz $\cA_i^{\bar ab} = a\psi J_i^{\bar ab}$, where $J_i^{\bar ab}$ are the generators of SU(2) in the $N$ dimensional representation, allows one to embed the scenario in an arbitrary rank gauge group.}
\be
\label{Aansatz}
\cA_i^a = \psi(t) a(t)\delta_i^a \ ,
\ee
the minisuperspace effective action for the axion $\cX$, scale factor $a$, and gauge field scale $\psi$ is
\be
\label{CrAct}
\cS = \int dt\,a^3 \Bigl[ -m_p^2\Bigl(\frac{\da}{a}\Bigr)^2 + \frac 32(\dot\psi+H\psi)^2 - \frac32 g^2\psi^4 + \half\dot\cX^2
	- \mu^4\Bigl(1+\cos\Bigl(\frac{\cX}{\fa}\Bigr)\Bigr) - \frac{3g\lambda\cX}{\fa}\psi^2(\dot\psi+H\psi) \Bigr] ~.
\ee
The scale factor $a(t)$ undergoes slow roll inflation for sufficiently large axion coupling $\lambda\gg 1$, small gauge coupling $g$, and appropriately tuned scale $\mu$ of the axion potential.  The key feature of the equations of motion in this parameter regime is that the axion velocity drives the growth of the gauge field, while the gauge field back-reacts to slow the motion of the axion sufficiently that the slow-roll conditions are satisfied, even when the axion potential isn't particularly flat.  In the action, the axion coupling to the gauge field is of the same order as the potential, both of which are much larger than the kinetic energy terms, when evaluated on the inflating trajectory; however, being a topological term, the $\cX \cF{\small\wedge}\cF$ term does not contribute to the stress tensor, and so the equation of state is potential dominated $P\approx-\rho$.

The plan for the remainder of this article is as follows: 
In section 2, we examine the dynamics of a charged particle in a magnetic field as a toy model, showing how the mini-superspace dynamics of chromo-natural inflation embed in this toy model, as well as how its dynamics exhibit the `magnetic drift' phenomenon.  
In section 3, we consider a sampling of the Chern-Simons terms in string theory, and focus on a particular model involving D7-branes.  This is particularly interesting in light of the role that such branes play in recent investigations of string model-building (see%
~\cite{Denef:2008wq} for a review).  
Section 4 contains a cursory look at how density perturbations are expected to arise in this context.
We conclude in section 5 with a summary discussion, comment on the stability and genericity of the mechanism, and propose a potential experimental signature.

%In order to make the presentation accessible to an audience of cosmologists as well as string theorists, at points the discussion will border on the pedantic to one or the other audience.  We hope the reader will bear with us.

%%%%%%%%%%%%%%%%%%%%%%%%%%%%%%%%%%%%%%%%%%%%%%%
%%%%%%%%%%%%%%%%%%%%%%%%%%%%%%%%%%%%%%%%%%%%%%%

\section{A toy model\label{toy}}

The underlying mechanism of chromo-natural inflation, and the intuition underlying its generalization, is elucidated by examining the dynamics of a charged particle.  The action
\be
\label{magact}
\cS =  \int dt\, \left[ \half\cG_{ab}(X) \dot X^a \dot X^b +\cA_a(X)\dot X^a - \cV(X) \right] 
\ee
generalizes the effective minisuperspace action \pref{CrAct}.  Consider the simple example of two-dimensional motion in a constant background magnetic field of magnitude $\lambda$ and harmonic potential well of frequency $\mu$.  If we add a small damping term, the equations of motion are
\bea
\ddot X + H\dot X + \mu^2 X &=& \lambda \dot Y \\
\ddot Y + H\dot Y + \mu^2 Y &=& -\lambda \dot X \ .
\eea
Ignoring for a moment the damping, there are two normal modes consisting of circular motion about the center of the well.  If the angular momentum is aligned with the magnetic field, the Lorentz force opposes the potential force; in the large field limit, `magnetic drift' balances the potential force against the magnetic force, so that $\mu^2 r_+ \approx \lambda v_+$.  
The orbital frequency of this mode is $\omega_+ = v_+/r_+ \approx \mu^2/\lambda$.  
The other normal mode has the angular momentum antialigned with the magnetic field, and in the large field limit the magnetic force balances against inertia, $v_-\approx \lambda r_-$.  
The orbital frequency is just the Larmor frequency $\omega_- = v_-/r_- \approx\lambda$. 
For equal energies, $\mu^2 r_+^2 \approx v_-^2$, the orbital radius of the fast mode is smaller than that of the slow mode by a factor $\mu/\lambda$.  The generic motion thus is a superposition of the two normal modes, and consists of an epicyclic motion around the center -- tight circles at the Larmor frequency superposed over a slow drift around the center at an angular frequency $\mu^2/\lambda$.  Putting in a small damping factor $H$, in the large field limit the fast mode damps at a rate $H$, while the slow mode damps more slowly, at a rate $H(\mu^2/\lambda^2)$.  Thus after some time, one is left only with the slow mode -- a gradual magnetic drift along an equipotential, with the orbit gradually spiraling down toward the center.  `Slow roll' is satisfied if the parameters are set in the hierarchy $\lam\gg H,\mu$.  

The minisuperspace action \pref{CrAct} of chromo-natural inflation is of the form \pref{magact}; to see this, let the field space in \pref{magact} have Lorentz signature, identify $T=a$, $X=\cX$ and $Y=a\psi$, and set
\bea
\cG_{ab} &=& \half \bma  -2m_p^2 T & 0 & 0 \\  0 & T^3 & 0 \\ 0 & 0 & 3T \ema  \nonumber\\
\cA_a &=& -3(g\lambda/\fa)\bigl( 0,0,XY^2) \\
\cV &=&\frac32 g^2Y^4/T + \mu^4T^3\bigl(1+\cos(X/\fa)\bigr)\ . \nonumber
\eea
Here, as usual, the Hamiltonian constraint arises when the particle wordline is coupled to an einbein $N(t)$ to enforce reparametrization invariance. This prevents the kinetic energy from being unbounded from below, despite the timelike signature of the scale
factor in the field space (DeWitt) metric that characterizes the minisuperspace dynamics~\pref{CrAct}.

To see the `magnetic drift' effect operating in chromo-natural inflation, consider the equations of motion in the slow-roll approximation, which after diagonalization of the velocities become
\bea
\label{Xdiag}
\Bigl(3H+\frac{g^2\lambda^2}{H\fa^2}\psi^4\Bigr)\dot\cX &=& \frac{\mu^4}{\fa}\sin(\cX/f) - \frac{g\lambda}{\fa} H\psi^3 + \frac{2g^3\lambda}{\fa H} \psi^5 \\
\label{psidiag}
\Bigl(3H+\frac{g^2\lambda^2}{H\fa^2}\psi^4\Bigr)\dot\psi &=& -2H^2\psi - 2g^2\psi^3 - \frac{g^2\lambda^2}{\fa^2}\psi^5 + \frac{g\lambda}{3H\fa^2}\psi^2\mu^4\sin(\cX/\fa) \ .
\eea
Of the two terms in the friction coefficient on the LHS, the first term represents Hubble friction, the second term arises from the magnetic drift force.
We want this `magnetic' force to dominate over Hubble friction, and therefore the coefficient of the Chern-Simons term should satisfy
\be
\label{CNIcondition}
g\lambda \gg \frac{\sqrt 3 \fa H}{\psi^2} \ .
\ee
The particular trajectory studied in%
~\cite{Adshead:2012kp}
sets the RHS of the second equation to zero, which is approximately solved in this regime by
\be
\label{psiequil}
\psi \approx \Bigl(\frac{\mu^4\sin(\cX/\fa)}{3g\lambda H}\Bigr)^{1/3}
\ee
and then the $\cX$ equation of motion determines a slow roll of the axion field governed by the effective potential gradient for $\cX$ determined by this locked-in value of $\psi$ on the RHS, and the larger-than-Hubble friction coefficient of $\dot\cX$ on the LHS; the motion can be made arbitrarily slow by dialing the axion coupling $g\lambda$~-- the strength of the analogue magnetic field~-- to be parametrically large.  This does not mean that we must have $g\lambda\gg1$.  Instead, this quantity merely needs to be larger than the RHS of~\pref{CNIcondition}.  
We also require $g^2\psi^4\ll \mu^4$, so that the axion potential dominates the gauge field energy, keeping the equation of state 
inflationary rather than radiation dominated. These two conditions, together with~\pref{psiequil}, require $\lambda\gg 1$.

This trajectory is not really special.  At large $\lambda$, both sides of the equation for $\dot\psi$ scale like $\lambda^2$; magnetic damping doesn't apply to its motion, and so the $\psi$ dynamics quickly evolves to this equilibrium point at a rate governed by Hubble damping (quite analogous to the toy model).  The lack of `magnetic damping' for $\psi$ can be traced to a Hubble friction term $\lambda H\psi^3$ in the Chern-Simons contribution to the $\cX$ equation of motion, which in turn leads to the term scaling as $\lambda^2$ on the RHS of \pref{psidiag}.  Conversely, the absence of a $\lambda^2$ term on the RHS of \pref{Xdiag} can be traced to the absence of such a Hubble friction term in the Chern-Simons contribution to the $\psi$ equation of motion.

Note that it is not necessary that the Chern-Simons coupling $g\lambda$ itself be much larger than one; we merely require~\pref{CNIcondition}.  On the trajectory~\pref{psiequil}, the condition~\pref{CNIcondition} is
\be
g\lambda \gg \frac{\fa^3\mu^2}{\mpl^5} \ ;
\ee
so we are allowed $g\lambda$ as small as we like, provided the scale $\mu$ of the potential is low enough relative to the Planck scale.  The parameter $\lambda$ serves as a useful indicator of the size of various contributions to the dynamics.  In what follows, we will employ an expansion in inverse powers of $\lambda$ as an approximation scheme.  This will be valid, even when $\lambda$ is not large (or even if it is dimensionful), because what matters is the hierarchy of scales between the Chern-Simons coupling, the Hubble scale, and the scale of the potential.

%%%%%%%%%%%%%%%%%%%%%%%%%%%%%%%%%%%%%%%%%%%%%%%
%%%%%%%%%%%%%%%%%%%%%%%%%%%%%%%%%%%%%%%%%%%%%%%

\section{String and M-theory realizations}

The key intuition gleaned from the toy model is that a magnetic-type coupling acts to oppose the potential force -- it keeps the fields from rapidly rolling to the minimum of the potential by diverting the motion in a direction orthogonal to the potential gradient.  The gauge field coupling $\cA(X)\!\cdot\!\dot X$ responsible for this effect is the simplest example of a Chern-Simons term, thus chromo-natural inflation should be but one example of a general mechanism of inflationary dynamics resulting from the interplay of an effective potential and a Chern-Simons coupling.  String theory exhibits a rich zoology of topological couplings, and therefore should prove a fertile source of such behavior.  Indeed, as we will see below, chromo-natural inflation embeds in string theory in a natural if not colorful way (for a review of axions in string theory, see for example%
~\cite{Svrcek:2006yi}). 

%%%%%%%%%%%%%%%%%%%%%%%%%%%%%%%%%%%%%%%%%%%%%%%

\subsection{An M-theory example (sort of)\label{Msect}}

Let us begin with M-theory.  The action of eleven-dimensional supergravity is
\be
\label{MAct}
\cS = \frac{1}{2\lpl^9} \int d^{11}x \left[ \sqrt{-g}R-\half G_4\wedge\star G_4-\frac16 C_3\wedge G_4\wedge G_4 \right] \ ,
\ee
where $\lpl$ is the eleven-dimensional Planck length, $C_3$ is a three-form antisymmetric tensor (AST) gauge field whose field strength is $G_4$; $C_3$ is a dimensionless periodic variable with period $2\pi$.
Consider a compactification to four dimensions with the warped FRW metric ansatz%
\footnote{Notation: $M,N,...$ are 10D indices; $\mu,\nu,...$ are indices along the 4D non-compact spacetime; $i,j,...$ non-compact spatial indices; $\alpha,\beta,...$ indices along the 7D compactification.}
\bea
\label{metric}
ds^2 &=& e^{-2A(x^\alpha)} h_{\mu\nu}dx^\mu dx^\nu + e^{A}g_{\alpha\beta}(x^\gamma)dx^\alpha dx^\beta\nonumber\\
&=& e^{-2A(x^\alpha)}\bigl(-N^2 dt^2+a(t)^2 (dx^i)^2\bigr) + e^{A}g_{\alpha\beta}(x^\gamma)dx^\alpha dx^\beta\ .
\eea
In what follows, we will suppress the effects of warping, and assume that we can write four-dimensional (or minisuperspace) effective actions, in which the warping is accounted for in the values of the parameters.

The reduction of the effective action to four dimensions, and further to minisuperspace, involves a decomposition of the various form fields into their various components along the macroscopic and compact directions.  For instance, the three-form field $C_3$ splits into a spatial three-form $C_{ijk}$, two-form $C_{ij\alpha}$, vector $C_{i\alpha\beta}$, and scalar $C_{\alpha\beta\gamma}$ as far as the macroscopic tensor properties are concerned.  The time components of the AST gauge potentials are Lagrange multipliers for Gauss constraints which we will largely ignore.  

Some of the modes $C_{\alpha\beta\gamma}$ may be light scalars in the four-dimensional effective action.  For instance, choose a harmonic 3-form $\omega_3$ in the compact manifold $\Sigma_7$ and set
\be
C_3 = \frac{\lpl^3}{2\pi}X(x^\mu) \omega_3(x^\alpha) \ ;
\ee 
classically $X$ is a massless scalar in the four-dimensional effective theory, because the underlying gauge symmetry $C_3\to C_3+d\Lambda_2$ forbids a potential for any of the components of $C_3$.  The effective 4d kinetic term for $X$ is 
\be
\frac{\gamma}{2} \int d^4x  \sqrt{h} h^{\mu\nu}\partial_\mu X\partial_\nu X 
\ee
where 
\bea
\gamma &=& \frac{1}{\lpl^9} \int_{\Sigma_7} \omega_3\wedge *\omega_3 
\eea
(here the Hodge star is with respect to the compactification seven-manifold, and $*\omega_3=\tilomega_4$). 
A variety of quantum effects can generate a potential $\cV(X)$; for instance, membrane instantons wrapping the dual three-cycle ${\Sigma}_3$ will induce a periodic potential for $X$%
~\cite{Harvey:1999as}.  Similarly, the kinetic energy for the spatial three-form $C_{ijk}$ reduces to
\be
\frac{V_7}{2\lpl^9} \int d^4x \frac1{\sqrt h} |\partial_0 C_{ijk}|^2 \ .
\ee

Let $G_4$ have a flux through the four-cycle $\tilsig_4$ dual to the 3-cycle $\Sigma_3$ conjugate to $\omega_3$
\be
\int_{\tilsig_4} G_4 = k\lpl^3 \ ;
\ee
the Chern-Simons form in~\pref{MAct} becomes
\be
\frac{k}{\lpl^3}\int d^4x X \partial_0 C_{ijk} \ .
\ee
The effective minisuperspace dynamics of the scale factor $a(t)$, the axion $X(t)$, and the scale of the spatial $C_3$
\be
\label{Cansatz}
C_{ijk} = a(t)^3 c(t)\, \epsilon_{ijk} \ ,
\ee 
specialize the four-dimensional effective field theory to spatially constant field configurations:
\bea
\label{msAct}
\cS_{\eff} = \int\! dt Na^3  \Bigl[ \frac{V_7}{N^2\lpl^9}\Bigl(-3\frac{{\dot a}^2}{a^2} + \frac{1}{2 a^6} \bigl(\partial_t(a^3 c)\bigr)^2
	 +  \gamma{\dot X}^2\Bigr) 
		- \cV(X)\Bigr] - \frac{k}{\lpl^3} X\partial_t(a^3 c) \ .
\eea
Once again we see dynamics with the structure \pref{magact}, and may imagine that for suitable choices of the parameters $\gamma$, $V_7/\lpl^7$, $k$, etc, the dynamics will lead to inflation, with magnetic drift dominating the evolution.

The minisuperspace equations of motion are%
\footnote{The variation of the metric should be made before substitutions of the sort~\pref{Cansatz}.}
\bea
\label{energyeq}
&&3\mpl^2 H^2 = \frac{\mpl^2}{4} (\dot c \!+\! 3Hc)^2
	+\frac{\mpl^2 \gamma}{2}\dot X^2 + \cV(X)  \\
\label{pressureeq}
&&\mpl^2 \bigl(2\dot H+3H^2\bigr) =  \frac{\mpl^2}{4} (\dot c \!+\! 3Hc)^2
	- \frac{\mpl^2\gamma}{2} \dot X^2 + \cV(X)  \\
\label{ceom}
&&\frac{\mpl^2}{2}\partial_t(\dot c+3Hc) - \delta \dot X = 0 \\
\label{Xeom}
&&\mpl^2\gamma \bigl({\ddot X} + 3H{\dot X} \bigr) + \cV'(X) +  \delta (\dot c+3Hc) = 0 \ .
\eea
Here the four-dimensional Planck scale $\mpl^2=V_7/\lpl^9$, and
$
\delta = k \mpl^3 \Bigl(\frac{\lpl^7}{V_7}\Bigr)^{3/2} 
$.
%Note that the four-form field strength generated by $c$ has equation of state $w=-1$, \ie\ it acts as a (negative) contribution to the cosmological constant.
%In four-dimensional terms, the axion potential is 
%\be
%\cV = \mpl^4 \;\alpha\Bigl(\frac{\lpl^7}{V_7}\Bigr)^2 \exp\!\left[-V_\Sigma/\lpl^3\right](1+\cos[X]) \equiv \varepsilon^4(1+\cos[X]) \ .
%\ee
%The usual fine-tuning of the final value of the vacuum energy has been introduced here.
%
The $c$ field equation of motion is readily integrated once
\be
\dot c+3Hc = \frac{2\delta}{\mpl^2}(X-X_0) \ ,
\ee
however, the integration constant $X_0$ is quantized%
~\cite{Beasley:2002db,Kaloper:2011jz}.  Briefly, the effective action
\be
\cS = \int d^4x \left( -\half G_4\wedge \star G_4 - \lambda X G_4 \right)
\ee
leads to the canonical momentum for the three-form potential
\be
\pi_C = G_4 - \lambda X = -\lambda X_0 \ .
\ee
In string/M-theory, the gauge group is compact, and so this momentum is quantized;
only discrete values of $X_0$ are allowed (different values correspond to changing the field by a flux quantum).
We thus find an equation for $X$ alone
\be
\label{Xeomeff}
\mpl^2\gamma \bigl({\ddot X} + 3H{\dot X} \bigr) + \cV'(X) +  \frac{2\delta^2}{\mpl^2}(X-X_0) = 0 
\ee
together with the Hamiltonian constraint
\be
3\mpl^2H^2 = \delta^2(X-X_0)^2+\frac{\mpl^2\gamma}{2}\dot X^2+\cV(X)
\ee
that determines $H$.

Despite the intuition gleaned from the toy model, this example does not generate slow-roll inflation; because the $c$ equation of motion can be integrated once, the velocity of the $c$ field is always tied to the displacement of the axion $X$ such that only one of the two normal modes analogous to those of the toy model is possible -- the one that rapidly decays and does not exhibit magnetic drift.

Now, it might seem that the intuition of the toy model is failing here, as the slow-roll conditions are not satisfied for the motion governed by \pref{Xeomeff}.  But actually, it works only too well -- the fast mode damps out as $X$ moves to the minimum of the effective potential
\be
\cV_{\rm eff} = \cV(X) + \frac{\delta^2}{\mpl^2}(X-X_0)^2
\ee
and the slow mode is {\it infinitely} slow -- there is no means for the dynamics to relax $X$ to the minimum of the original potential $\cV(X)$, rather it gets stuck at the minimum of this effective potential, which for large $\delta$ is fixed at a point set to a quantized value by the $c$ field dynamics (modulo nucleation of membranes that can shift $X_0$ by discrete amounts%
\cite{Bousso:2000xa}).

The quadratic term in the effective potential has the same source as in the Schwinger model%
~\cite{Coleman:1976uz,Beasley:2002db,Kaloper:2011jz}.
When the rank of an AST potential equals the spatial dimension, there is no magnetic field, only an electric field; in the analogue of 
%the 
Amp\`ere's law, there is no magnetic field for the the electric field energy to oscillate into, and so the electric field dynamics is rigidly related to that of the charged sources.  In order to bypass this obstruction, we can try using AST fields of lower rank, as for example the one-form fields of chromo-natural inflation.  Such terms arise when we add branes to the mix of ingredients; let us now turn to some string theory examples.

%%%%%%%%%%%%%%%%%%%%%%%%%%%%%%%%%%%%%%%%%%%%%%%

\subsection{String theory}

A wide variety of examples of EM-flation arises from compactifications involving space-filling D-branes.  Our starting point is the ten-dimensional supergravity action; for type IIB, this reads
% see for example [ref???polchinski] 
\bea
\label{IIBAct}
\cS_{\rm IIB} &=& \frac{2\pi}{\lstr^8}\int \sqrt{-g}\Bigl[ e^{-2\Phi}\Bigl(R+4(\partial\Phi)^2-\half |H_3|^2 \Bigr) \\
	& & \hskip 3cm - \half\Bigl( |F_1|^2+|\tilde F_3|^2+\half |\tilde F_5|^2 \Bigr) - \half C_4\wedge H_3 \wedge F_3 \Bigr]  \nonumber
\eea
while for type IIA one has
\bea
\label{IIAAct}
\cS_{\rm IIA} &=& \frac{2\pi}{\lstr^8}\int \sqrt{-g}\Bigl[ e^{-2\Phi}\Bigl(R+4(\partial\Phi)^2-\half |H_3|^2 \Bigr) \\
	& & \hskip 3cm - \half\Bigl( |F_2|^2+|\tilde F_4|^2 \Bigr) - \half B_2\wedge F_4 \wedge F_4 \Bigr] \ . \nonumber
\eea
Here $F_{n+1}$ are the $(n\!+\!1)$-form field strengths $F_{n+1}=dC_n$ of $n$-form antisymmetric tensor (AST) gauge fields $C_n$, the three form field strength $H_3$ arises from a two-form potential $B_2$, and
\bea
\tilde F_3 &=& F_3 - C_0\wedge H_3 \ , \nonumber \\
\tilde F_5 &=& F_5 - \frac12 C_2\wedge H_3 + \frac12 B_2\wedge F_3 \\
\tilde F_4 &=& F_4 - C_1\wedge H_3 \ . \nonumber 
\eea
These supergravity fields are coupled to Dp-brane sources through their worldvolume action
\bea
\label{DpAct}
\cS_{\rm Dp} &=& -\frac{2\pi}{\lstr^{p+1}} \int d^{p+1}\xi\, \Tr\Bigl[ e^{-\Phi}\bigl[-\det(G_{ab}+B_{ab}+2\pi\alpha' \cF_{ab})\bigr]^{1/2} \\
	& & \hskip 4cm - \Bigl(\sum_n C_n\Bigr)\wedge \exp[B_2+2\pi\alpha' \cF_2] \Bigr] \ ;\nonumber
\eea
the bulk supergravity fields $G_{MN}$, $B_{MN}$, $\Phi$ and $C_{M_1...M_n}$ are pulled back to the brane using the worldvolume parametrization $X^M(\xi^a)$; $\cF_{ab}$ is the field strength of a gauge potential $\cA_a$ intrinsic to the brane, and $2\pi\alpha'=\lstr^2/(2\pi)$.  
Note that the scalar fields $X $ are not canonically normalized, and carry the dimensions of length since they parametrize motion in the extra dimensions.  For simplicity, we have suppressed additional couplings in the Chern-Simons term on the second line involving curvatures of the brane worldvolume and its normal bundle%
~\cite{Minasian:1997mm,Cheung:1997az}, as well as commutators of the normal bundle scalars $X $%
~\cite{Myers:1999ps}.  The curvature couplings enter various equations of motion for the antisymmetric tensor field (in particular charge conservation equations), while the additional scalar terms provide additional sources of examples.

In the effective dynamics, there will be additional contributions to the effective action from various matter fields living on brane intersections, potentials contributed by gauge field and $Dp$-brane instantons, supersymmetry breaking, fluxes, etc.  While these may be important for model building, all we will be concerned with is the general form and scale of the effective potential.

%%%%%%%%%%%%%%%%%%%%%%%%%%%%%%%%%%%%%%%%%%%%%%%
%%%%%%%%%%%%%%%%%%%%%%%%%%%%%%%%%%%%%%%%%%%%%%%

\subsection{Effects of the D-brane Chern-Simons term\label{CSgeneralities}}

Consider the Chern-Simons term on a {\it Dp}-brane that fills non-compact space and spans $(p-3)$ directions in the compactification manifold $\Sigma_6$.  It will involve the AST fields $C_n$, which under reduction to four dimensions lead to spatial form fields of ranks from zero to three; $B_2$, whose dynamical components consists of spatial two-forms $B_{ij}$, vectors $B_{i\alpha}$, and scalars $B_{\alpha\beta}$; gauge fields $\cA$ that similarly reduce to vectors and scalars in 4d; and the scalars $X $ that specify the brane's location and orientation in the compactification manifold.  

In the warped FRW metric \pref{metric}, an appropriate ansatz for these various fields takes the form
\be
\label{fieldansatz}
\Phi_{i_1\dots i_k \alpha_1\dots \alpha_m} = \sum_p a(t)^k \phi^{(p)}_{i_1\dots i_k}(t) \omega^{(p)}_{\alpha_1\dots \alpha_m} (x^\beta)
\ee
where the $\omega^{(p)}$ are a basis of cohomology in $\Sigma_6$.%
\footnote{This is a slight oversimplification due to the possibility of evolution of the fields within $\Sigma_6$ as the brane moves, due to back-reaction.}
Upon integration over $\Sigma_6$, the structure of the Chern-Simons coupling in~\pref{DpAct} is a collection of terms of the form
\be
\label{CSstruct}
\lambda_P\, a^{3-m} \phi_{(p_1)} \dots \phi_{(p_{k})} \partial_t(a^{m}\phi_{(p_{k+1})}\dots\phi_{(p_n)}) \ ,
\ee
with $m=1$ for terms involving brane gauge field strengths $\cF_{ti}$, and $m=0$ from terms involving field strengths $\cF_{t\alpha}$ or brane velocities $\dot X $.%
\footnote{In principle there are terms as well from derivatives of the $\omega^{(p)}$ induced by the brane motion $\dot X $; we will ignore these effects in what follows, however they will lead to additional terms of the same general form as~\pref{CSstruct}.}
Let us focus on a single such term, and suppress the multi-index $\lambda_P\equiv\lambda$.
In the equations of motion that follow from \pref{IIBAct}, or \pref{IIAAct}, and \pref{DpAct}, the slow-roll approximation \pref{slowroll} ignores the accelerations of the fields in favor of Hubble friction; for a spatial $m$-form,%
\footnote{Possible kinetic mixing among the fields will be ignored for simplicity; including it does not qualitatively change the results.}
the kinetic term reduces to
\be
\label{slowrollkinetics}
a^m\partial_t\bigl(a^{3-2m}\partial_t(a^m\phi)\bigr)\approx a^3\bigl(3H\dot\phi + m(3-m)H^2\phi\bigr) \ .
\ee
The schematic structure of the slow-roll equations of motion is thus
\be
\label{eomeff}
\cM_{pq}\dot\phi_{(q)} ={ \bf F}_{p}(\phi)
\ee
where the matrix $\cM$ has the structure
\be
\label{Mstruct}
\cM = \left(\begin{matrix} 3H \cI_{k} & \lambda \cP \\ -\lambda\cP^t & 3H \cI_{n-k}\end{matrix}\right)
\ee
with $\cI_k$ the rank $k$ identity matrix and $\cP$ the rectangular matrix 
\be
\cP_{ij}=\frac{\partial^2(\prod \phi)}{\partial\phi_{(p_i)}\partial\phi_{(p_j)}}\ , 
\ee
with $i=1,..., k$; $j=k\!+\!1,...,n$; and $\prod \phi$ is the product of all the fields appearing in the Chern-Simons term~\pref{CSstruct}.  The force term on the RHS of~\pref{eomeff} includes all contributions from potentials of effective scalars, as well as the terms proportional to $H^2$ from the kinetic terms~\pref{slowrollkinetics}, and non-derivative terms proportional to $\lambda H$ from the Chern-Simons term~\pref{CSstruct}.  

Diagonalizing the velocities leads to 
\be
\label{diagvel}
\det[\cM]\dot\phi_{(p)} = (\det[\cM]\cM^{-1})^{pq}{\bf F}_q(\phi)\ .
\ee
To this end, note that if the Chern-Simons interaction is linear in each of the $n$ fields appearing, the matrix $\cM$ can be written%
\footnote{The generalization to Chern-Simons terms of more general functional form is straightforward.}
\be
\hat \cM = \frac{1}{3H}\cM = \One + z\Bigl(\ket{\alpha}\bra{\beta} - \ket{\beta}\bra{\alpha}\Bigr)
\ee
where $z=({\lambda}/{3H})\prod_{i=1}^n\phi_{(p_i)}$ and $\ket\alpha$, $\ket\beta$ are vectors in the space spanned by the $\phi_{(p_i)}$ with components
\bea
\vev{i|\alpha} &=& \frac{1}{\phi_{(p_i)}}~~,\quad i=1,\dots, k \ ;\quad \vev{i|\alpha}=0~~,\quad i=k+1,\dots,n \nonumber\\
\vev{j|\beta} &=& \frac{1}{\phi_{(p_j)} }~~,\quad j=k+1,\dots, n \ ;\quad \vev{j|\beta}=0~~,\quad j=1,\dots,k
\eea
One then has $\det[\hat\cM]=1+z^2\vev{\alpha|\alpha}\vev{\beta|\beta}$, and 
\be
\det[\hat\cM]\hat \cM^{-1} = \det[\hat\cM]\One-z\Bigl(\ket\alpha\bra\beta-\ket\beta\bra\alpha\Bigr)-z^2\Bigl(\ket\alpha\vev{\beta|\beta}\bra\alpha+\ket\beta\vev{\alpha|\alpha}\bra\beta\Bigr)
\ee
The force vector ${\bf F}$ in~\pref{eomeff} has the structure 
\be
\label{forcevec}
{\bf F}=zH\Bigl(m\ket\alpha - (3-m)\ket\beta\Bigr) + \bF
\ee
where $\bF$ is independent of $\lambda$.  
Thus, in the space orthogonal to the two vectors $\ket\alpha$, $\ket\beta$, there is no `magnetic damping' effect, velocities are of order $\lambda^0$, and fields will evolve quickly to some quasi-static point of equilibrium with roughly Hubble-size damping.%
\footnote{In principle, we should re-solve the equations of motion for these directions in field space, without making the slow-roll approximation, since slow roll does not apply to them.  Instead, we will ignore this complication and assume that these fields have evolved to a quasi-static equilibrium value.}
 
Within the two-dimensional subspace spanned by $\{\ket\alpha,\ket\beta\}$, one has
\bea
\det[\hat\cM] \hat\cM^{-1}\ket\alpha &=& \ket\alpha+z\ket\beta\vev{\alpha|\alpha} \nonumber \\
\det[\hat\cM] \hat\cM^{-1}\ket\beta &=& \ket\beta-z\ket\alpha\vev{\beta|\beta}
\eea
and there is indeed a `magnetic damping' effect -- the $\det[\cM]$ coefficient on the LHS of~\pref{diagvel} is order $\lambda^2$, while the matrix $\det[\hat\cM]\hat\cM^{-1}$ restricted to the subspace is order $\lambda$.  However, in one of these two directions the force vector is of order $\lambda$, and combined with a contribution of order $\lambda$ from $\det[\hat\cM] \hat\cM^{-1}$, the force in this direction is also or order $\lambda^2$ and competes with the damping.  The direction
\be
\ket{\chi} =m\ket\alpha-(3-m)\ket\beta
\ee
is orthogonal to the leading $\lambda^2$ force in~\pref{diagvel} (in \pref{Xdiag}-\pref{psidiag}, this was the $\cX$ direction); in this direction, the damping coefficient $\det\cM$ is effective in slowing the motion to a velocity that is parametrically of order $1/\lambda$.
% times a function of the quasi-static equilibrium field values obtained by solving the remaining equations for the undamped directions.
The direction $\ket\phi$ orthogonal to $\ket\chi$ in the space spanned by $\ket\alpha$, $\ket\beta$, 
\be
\ket\phi = (3-m)\ket\alpha\vev{\beta|\beta}+m\ket\beta\vev{\alpha|\alpha} \ ,
\ee
as well as all the directions orthogonal to $\ket\alpha$, $\ket\beta$, the velocities are not parametrically suppressed in $\lambda$, since the force is of the same order in $\lambda$ as the damping coefficient.

The effective field parametrizing the $\ket\chi$ direction is our candidate inflaton.  The values of all the other fields are found from the solution to the field equations as a function of this slow variable; this solution will bring additional dependence on $\lambda$ into the effective equation for the inflaton motion (as in the example~\pref{psiequil}), thus an explicit expression for the number of e-foldings of inflation as a function of $\lambda$ (and other parameters in the effective action) will depend on a case-by-case analysis.

%%%%%%%%%%%%%%%%%%%%%%%%%%%%%%%%%%%%%%%%%%%%%%%
%%%%%%%%%%%%%%%%%%%%%%%%%%%%%%%%%%%%%%%%%%%%%%%

\subsection{D7-brane example\label{D7sect}}

Consider a type IIB compactification to four dimensions with $N$ D7-branes.  
The motion of the branes is not frozen by space-filling form-field dynamics; the D7-brane locations are pseudo-moduli, the locations of fiber degenerations in a description of the compactification $\Sigma_6$ as the base of an elliptically fibered Calabi-Yau fourfold $Y_4$ in F-theory%
~\cite{Vafa:1996xn} (for a review, see%
~\cite{Denef:2008wq}).
%A tuneable parameter is afforded by the Chern-Simons coupling
%\be
%C_{3\alpha\beta\gamma}B_{12} \dot X \wedge\cF\ ,
%\ee
%where we wrap the $N$ D7's along a four-cycle $S$ threaded by a gauge field of first Chern class $c_1=K$.  

We expect that the four-form potential that D7-branes couple to directly cannot generate inflationary dynamics through the analogue of magnetic damping, because space-filling form fields are subject to the mechanism of section~\ref{Msect}.  However, there are additional couplings in the Chern-Simons term, and we might hope that one of them can provide the effect of interest.

What other terms might yield inflation?  Consider the contribution to the D7-brane Chern-Simons term that couples to the brane velocity as
\be
C_{i\alpha_1\alpha_2\alpha_3}\wedge B_{jk}\wedge (2\pi\alpha'\cF)_{\alpha_4\alpha_5}\dot X^{\alpha_1} \ ;
\ee
and adopt the ansatz \pref{fieldansatz} (so that $B_{12}=a^2b$, $C_{3\alpha\beta\gamma}=a\,c\,\omega_{\alpha\beta\gamma}$; the expectation values for $B$, $C$ break rotational symmetry and we choose this orientation of the vevs), and take $\cF$ to have $K$ units of flux through a two-cycle wrapped by the brane.%
\footnote{Note that, for this term to be active during inflation, both $B_2$ and $C_4$ must be lighter than the Hubble scale; however, in a typical compactification breaking to $N=1$, one of these fields gets a mass at least of order the compactification scale.  For this realization of EM-flation to be viable, one would need the compactification to preserve $N=2$ supersymmetry, with $N=2$ in the bulk broken at a scale somewhat below the Hubble scale.}
The effective action reads (we henceforth suppress the internal indices)
\be
\label{D7msAct}
\cS_{\eff} = \int\! dt Na^3  \Bigl[ \frac{1}{N^2}\Bigl(-3\mpl^2\frac{{\dot a}^2}{a^2} + \frac{\gamma_b}{2 a^4} \bigl(\partial_t(a^2 b)\bigr)^2
	 + \frac{\gamma_c}{2a^2 } \bigl(\partial_t(a c)\bigr)^2
	 +  \frac{\gamma_X}{2}{\dot X}^2\Bigr) 
		- \cV(X)\Bigr] - \lambda a^3bc \dot X  \ .
\ee
The Chern-Simons term takes the general form~\pref{CSstruct}, and the equations of motion in the slow-roll approximation \pref{slowrollkinetics} become
\bea
\label{BCXeoms}
\gamma_b (3H\dot b + 2H^2 b) &=& \lambda c  \dot X  \nonumber\\
\gamma_c (3H\dot c  + 2H^2 c ) &=& \lambda b \dot X \\
\gamma_X (3H\dot X) +\cV'(X) &=& - \lambda ( c \dot b+b\dot c +3Hbc ) \nonumber
\eea
Diagonalizing the velocities as in \pref{diagvel}, one finds that the $X$ equation of motion boils down to 
\be
\label{XeomBCX}
[9 \gamma_b \gamma_c \gamma_X H^2 + \lambda^2( \gamma_b b^2  + \gamma_c c^2 )] \dot X = 
	-\gamma_b\gamma_c H (3 \cV' + 5H \lambda b c  )
\ee
while the force terms on the RHS of the $b$, $c $ equations of motion 
\be
\label{CeomBCX}
\left[9 \gamma_b \gamma_c \gamma_X H^2 + \lambda^2( \gamma_b b^2  + \gamma_c c^2 )\right] \dot c  =
- \Bigl[ \gamma_b \lambda b  \cV'  + c \Bigl(\lambda^2 H\bigl(\frac23\gamma_c c^2+\frac73\gamma_b b^2\bigr)+6\gamma_b\gamma_c\gamma_X H^3\Bigr) \Bigr]  
\ee
\be
\label{BeomBCX}
\left[9 \gamma_b \gamma_c \gamma_X H^2 + \lambda^2(\gamma_b b^2+\gamma_c c^2)\right]  \dot b =
- \Bigl[ \gamma_c \lambda c \cV'  + b\Bigl(\lambda^2 H\bigl(\frac73\gamma_c c^2+\frac23\gamma_b b^2\bigr)+6\gamma_b\gamma_c\gamma_X H^3\Bigr) \Bigr]
\ee
are of order $\lambda^2$, the same as the damping coefficient on the LHS, in accord with the general analysis of the previous section.  In other words, $b$ and $c $ quickly evolve to a nonzero local minimum,%
\footnote{There is also of course the trivial solution with $b=c =0$, which one might think leads to ordinary D-brane inflation if one could solve the attendant problems mentioned in the introduction.  Below, we will argue that this trajectory is unstable.}
obtained by setting $\dot b=\dot c =0$; and then the inflaton $X $ slowly rolls down its effective potential.  Plugging in the quasi-equilibrium values of $b$, $c $ at large $\lambda$, 
\bea
b^* &\approx& \left(\frac{\cV'}{3\lambda H}\sqrt{\frac{\gamma_c}{\gamma_b}} \;\right)^{1/2} \nonumber\\
\label{bcattractor}
c ^* &\approx& \left(\frac{\cV'}{3\lambda H}\sqrt{\frac{\gamma_b}{\gamma_c}} \;\right)^{1/2}
\eea
one finds that the $X $ velocity scales as
\be
\label{Xvel}
\dot X  \approx 
\frac{2\sqrt{\gamma_b\gamma_c}\, H^2}{\lambda} \ .
\ee
There is a second attractor point at $b=-b^*$, $c =-c ^*$ with the same solution for $\dot X $.
The $\lambda^{-1}$ scaling of $\dot X $ should not be too surprising, the structure of the equations of motion is quite similar to chromo-natural inflation -- the form fields $b$, $c $ play the same role as the chromo-natural gauge potential $\cA$, while $X $ plays the same role as the chromo-natural axion $\cX$.  

One aspect that we have until now ignored is that the specific Chern-Simons coupling chosen is not at all isotropic in the inflating directions, nevertheless we have assumed that the metric inflates isotropically.  In fact, given this anisotropy, we should generalize the metric ansatz to allow independent scale factors $a_1$, $a_2$, $a_3$ for the three inflating directions.   The equations of motion for the $b$, $c $ fields become
\bea 
\gamma_b H_{tot} \dot b+\gamma_bH_3(H_1+H_2)b-\lambda c \dot X  &=& 0 \nonumber \\
\gamma_c H_{tot} \dot c +\gamma_cH_3(H_1+H_2)c -\lambda b\dot X  &=& 0 
\eea
where $H_{tot}=H_1+H_2+H_3$.  Multiply the first equation by $b$ and the second by $c $ and subtract to find 
\be
H_{tot}\partial_t(\gamma_b b^2-\gamma_c c^2)+H_3(H_1+H_2)(\gamma_b b^2-\gamma_c c^2) = 0 \ .
\ee
This tells us that $\gamma_b b^2=\gamma_c c^2 $ up to terms that die off exponentially in time.

Now consider the pressure difference equations, the differences of the equations of motion for the $a_i$.  The nontrivial ones are of the form
\be
\label{pressurediff}
0 = \partial_t(H_3-H_2)+H_{tot}(H_3-H_2)+\frac{2\gamma_b}{\mpl^2}(\dot b+(H_1+H_2)b)^2-\frac{2\gamma_c}{\mpl^2}(\dot c +H_3 c )^2 
\ee
Plug in the equations of motion, and use the fixed point $\gamma_b b^2=\gamma_c c^2 $ to find
\be
\label{anisotropy}
0 = \partial_t(H_3-H_2)+H_{tot}(H_3-H_2) -  \frac{2\gamma_b b^2}{\mpl^2H_{tot}^2}\Bigl(H_3^2+(H_1+H_2)^2+\frac{2\lambda}{\sqrt{\gamma_b\gamma_c}}\dot X\Bigr)\left( (H_1+H_2)^2-H_3^2\right)
\ee
up to exponentially decaying terms.   The last term, which sources $(H_3-H_2)$, is of order $\sqrt{\gamma_b\gamma_c}\; \cV'/(H_{tot}\lambda)$ times a quadratic function of the $H_i$, and so the anisotropy in Hubble expansion rates is formally of order $1/\lambda$.  We will largely ignore this issue for now, but will return to it in the discussion section below.

It remains to check that one can engineer a large `magnetic' coupling $\lambda$ in the effective equations of motion.  For $N$ D7-branes with $K$ units of gauge flux, dominance of the magnetic drift term in the damping coefficient in the equations of motion~\pref{XeomBCX}-\pref{BeomBCX} requires
\be
\sqrt{\gamma_b\gamma_c}\; \gamma_X H^3 \ll \cV' \lambda\ ;
\ee
in our conventions, we have
\bea
\gamma_b &=& \frac{V_6}{\gstr^2\lstr^8} \equiv \mpl^2 \nonumber\\
\gamma_c &=& \frac{V_6}{\lstr^8} = \gstr^2\mpl^2  \\
\gamma_X &=& \frac{N}{\gstr\lstr^4} \nonumber\\
\lambda &=& \frac{K}{\lstr^4} \nonumber
\eea
where $\mpl$ is the effective four-dimensional Planck scale.  The potential for brane motion has a characteristic scale $\mu$ and varies over the size $R$ of the compactification manifold (so for a generic compactification $V_6\sim R^6$)
\be
\cV = \mu^4 N f(X/R)\ .
\ee
With $3\mpl^2 H^2 = \cV$, the condition for dominance by the Chern-Simons interaction becomes
\be
\frac{N^{1/2}}{K}\; \frac{\mu^2 R}{\mpl} \frac{f^{3/2}}{f'} \ll 1 \ ,
\ee
which can be satisfied by making the scale of the effective potential small enough relative to the geometric mean of the compactification scale and the four-dimensional Planck scale, or by making the flux quantum $K$ large.%
\footnote{The parameter $\lambda$ is generically dimensionful; more precisely, when we say that we want to take $\lambda$ to be large, we mean that its scale is typically the string or Planck scale, and we wish to take the other scales in the problem (such as the scale $\mu$ of the potential, and the compactification scale) to be small relative to this scale.}
The velocity~\pref{Xvel} is then
\be
\label{Xdot}
\dot X  \approx \frac{2N}{3K}\gstr\lstr^4 \mu^4 f(X/R) = \frac{2\mu^4R^3}{3\mpl} \frac{N}{K} f(X/R) \ .
\ee
Again can be made small by making the scale of the effective potential for brane motion low enough, or $K$ large.  Taking the range of inflationary motion to be of order~$R$,%
\footnote{Note that while the radius of compactification is larger than the ten-dimensional Planck scale, this does not mean that the range of motion of the inflaton is greater than the four-dimensional effective Planck scale.  The normalized scalar deformation relative to the four-dimensional Planck scale is $(\sqrt{\gamma_X}\Delta X)/\mpl \sim \gstr^{1/2}(\lstr/R)^2<1$.}
the number of e-foldings of inflation is estimated to be
\be
\label{efolds}
N_e \sim H\Delta t \sim \frac{K R\, \vev{f^{-1/2}}}{N^{1/2}\mu^2 V_6^{1/2}} \sim \frac{K}{N^{1/2}\mu^2 R^2} \ .
\ee
This is again large when $\mu R\ll 1$, and/or $K\gg 1$.  The slow-roll parameters~\pref{slowroll} are calculated as%
\footnote{Note that standard expressions such as $\eta_H=-\ddot X/(H\dot X)$ are derived from~\pref{slowroll} using the equations of motions of single scalar field inflation, and do not apply to our model.} 
\bea
\label{etaeps}
\epsilon_H &=& -\frac{\dot H}{H^2} = -\frac{\sqrt 3\,\mpl}{2} \; \frac{\cV' }{\cV^{3/2}} \; \dot X\nonumber\\
\eta_H +\half\epsilon_H &=& -\frac{\ddot H}{2H\dot H} -\frac{\dot H}{2H^2}
\sim -\frac{\sqrt 3\,\mpl}{2} \;\frac{\cV''}{\cV' \cV^{1/2}} \; \dot X \ .
\eea
Note that, since both $\epsilon_H$, $\eta_H$ are proportional to $\dot X$, which is of order $1/\lambda$, these slow roll parameters can be made small.  Using the estimate~\pref{Xdot}, we have
\bea
\epsilon_H & \sim& 
-\Bigl(\frac{\mu^2 R^2N^{1/2}}{K} \frac{f'}{f^{1/2}} \Bigr) \nonumber\\
\eta_H +\half\epsilon_H &\sim& 
  - \Bigl(\frac{\mu^2R^2N^{1/2}}{K}\frac{f'' f^{1/2}}{f'}\Bigr) 
\sim \epsilon_H \sim \frac{1}{N_e} \ .
\eea
From~\pref{anisotropy}, the anisotropy of the expansion is estimated to be 
\be
\label{deltaH}
\frac{H_3-H_2}{H_{tot}} \sim - \frac{\mu^2R^2N^{1/2}}{K}\frac{f'}{f^{1/2}} \sim \epsilon_H \ .
\ee

%%%%%%%%%%%%%%%%%%%%%%%%%%%%%%%%%%%%%%%%%%%%%%%
%%%%%%%%%%%%%%%%%%%%%%%%%%%%%%%%%%%%%%%%%%%%%%%

\subsection{Chromo-natural inflation on branes}

Chromo-natural inflation embeds easily on D-branes.  Here we can take for instance the part of the D3-brane Chern-Simons interaction
\be
\label{axion}
C_0\Tr[\cF\wedge\cF] \ .
\ee
The minisuperspace effective action becomes
\be
\label{CND3Act}
\cS_{\eff} = \int\! dt Na^3  \Bigl[ \frac{1}{N^2}\Bigl(-3\mpl^2\frac{{\dot a}^2}{a^2} + 
\gamma_\cA \frac{\Tr\![(\dot \cA)^2]}{a^2}  +\gamma_c \dot c^2\Bigr) + \mu^4 \hat\cV(c)+ 
\gamma_\cA \frac{\Tr\!([\cA,\cA]^2)}{a^4}\Bigr]  +
\kappa\, c\, \Tr\!(\dot \cA[\cA,\cA])  \ ,
\ee
with $\gamma_\cA=1/(\gstr (2\pi)^2)$, $\gamma_c=\gstr^2\mpl^2=R^6/\lstr^8$, and $\kappa=1/(2\pi)^2$.  Note that $c$ is a dimensionless variable of unit period.  The only change for D7-branes is that $\kappa$ gets an integer factor from the compactification, and $\gamma_\cA$ picks up a factor of the volume $V_S/\lstr^4$ of the 4-cycle wrapped by the D7.
The $SU(N)$ embedding discussed in the footnote around equation~\pref{Aansatz} leads to the following field identifications:
\bea
C_0 &=& \frac{\cX}{\fa} \nonumber\\
\cA_i &=& \frac{a\psi J_i}{(\gamma_\cA \nu)^{1/2}}
\eea
(with $\Tr[J_i J_j] = \nu\delta_{ij}$ where $\nu=N(N^2-1)/12$), and parameter identifications:
\bea
\label{CNIparamsMap}
g &=& \frac{1}{(\gamma_\cA\nu)^{1/2}} \nonumber\\
\fa &=& \sqrt{\gamma_c} \\
\lambda &=& \frac{\kappa}{\gamma_\cA} \ .\nonumber
\eea
Note that on D3-branes, one has $\lambda\sim\gstr<1$, so the conditions for chromo-natural inflation are not met.  This problem can be circumvented by passing to \eg\ D7-branes and inducing~\pref{axion} using gauge flux along the internal directions of the D7-brane worldvolume.

A tuneable parameter is afforded by the Chern-Simons coupling
\be
C_0 \Bigl(\frac{1}{24}\Tr[\cF\wedge\cF\wedge\cF\wedge\cF] + \half\Tr[\cF\wedge\cF]\hat\chi(R)/24\Bigr)\ ,
\ee
where we wrap the $N$ D7's along a four-cycle $S$ of Euler number%
\footnote{$\chi(S)=\int_S \hat\chi(R)$ is the integral of the Euler density $\hat\chi(R)$, one of the curvature couplings we suppressed in writing the Chern-Simons coupling in~\pref{DpAct}.}
$\chi(S)$ threaded by a gauge field of instanton number $c_2=K$.  We can then try to tune $(K+\chi(S)/24)$ to be large, subject to Gauss law and tadpole cancellation constraints.  
The integrated flux is subject to a tadpole cancellation condition,
\be
\int_{S}\half \Tr[\cF\wedge\cF] +\frac{1}{24}\Bigl(\chi(D7)+4\chi(O7)\Bigr) =  \frac{1}{\lstr^4}\int_{\Sigma_6} F_3\wedge H_3 + 2N_{\rm D3} 
\ee
where $S$ is the four-dimensional cycle in $\Sigma_6$ wrapped by the D7, and $\chi$ is the Euler character of the cycle $S$ wrapped by any D7-branes (similarly for O7 orientifold seven-planes).  As usual, one can dissolve free-standing D3-branes into gauge flux on the D7, so if $N_{\rm D3}$ can be large, then one may have a substantial freedom in tuning the Chern-Simons coupling for inflation.  This tadpole cancellation condition in turn descends from a corresponding condition in the F-theory fourfold
\be
\frac{\chi(Y_4)}{24} = \half \int_{Y_4} G_4\wedge G_4 + N_{\rm D3} 
\ee
and the LHS in some recent examples used for GUT model-building%
~\cite{Denef:2008wq,Marsano:2009ym,Collinucci:2008pf,Knapp:2011wk} 
is of order tens to hundreds.
The parameters~\pref{CNIparamsMap} are now determined by
\bea
\gamma_\cA &=& \frac{V_S}{(2\pi)^2\gstr\lstr^4} \nonumber\\
\kappa &=& \frac{K+\chi(S)/24}{(2\pi)^2}
\eea
and the Chern-Simons interaction is made more dominant either by making the flux $K$ large or the volume $V_S$ of the four-cycle $S$ small, or both.

In the usual static approximation~\pref{psiequil}, the condition $g^2\lambda^2\psi^4 \gg 3H^2 \fa^2 $ becomes
\be
g\lambda \mu^8 \hat\cV^2 \gg 3^{1/2}H^5 \fa^3 \ ,
\ee
and with the above substitutions one finds the condition
\be
K\frac{ 3^2\cdot2\pi}{(\gstr^3\nu \hat\cV )^{1/2} }\frac{\mpl^2}{\mu^2} \gg 1 \ .
\ee
This condition is satisfied when $\mu \ll \mpl$ and $\gstr N$ is small.

The axion velocity can be determined from the equations of motion~\pref{Xdiag}, \pref{psidiag}.  The analysis of%
~\cite{Adshead:2012kp,Adshead:2012qe}
showed that the number of e-folds of inflation is maximized for $g^2\psi^2\sim H^2$; with this specialization, one finds a simple expression for the axion velocity
\be
\dot C_0 = \frac{\dot \cX}{\fa} \approx\frac43  \frac{H\mu^4 \hat\cV'}{g^2\lambda^2\psi^4} 
\ee
which scales as $\lambda^{-2/3}$ on the quasi-static solution~\pref{psiequil}:
\be
\frac{\dot C_0}{H} = \frac{\dot \cX}{\fa H} \approx \frac{4}{3^{1/3}(2\pi)^{2/3}}\frac{\nu^{1/3}V_S}{\gstr K^{2/3}\lstr^4}\left(\frac{\mu^4}{\mpl^4} \;\frac{\hat\cV^2}{\hat\cV'}\right)^{1/3} \ .
\ee
Inserting this result into the slow roll parameters~\pref{etaeps}, one finds
\bea
\epsilon_H &\sim& \frac{NV_S}{\gstr K^{2/3}\lstr^4} \Bigl(\frac{\mu}{\mpl} \Bigr)^{4/3} \;\frac{(\hat\cV')^{2/3}}{\hat\cV^{1/3}} \nonumber\\
\eta_H+\half\epsilon_H &\sim&  \frac{NV_S}{\gstr  K^{2/3}\lstr^4} \Bigl(\frac{\mu}{\mpl} \Bigr)^{4/3} \;\frac{\hat\cV'' \hat\cV^{2/3}}{(\hat\cV')^{4/3}} \ .
\eea
One can again analyze the difference of the pressures, with the result in this case that differences in the Hubble expansion rates damp exponentially in time.  This is not surprising, due to the symmetry among the spatial directions.   

%%%%%%%%%%%%%%%%%%%%%%%%%%%%%%%%%%%%%%%%%%%%%%%
%%%%%%%%%%%%%%%%%%%%%%%%%%%%%%%%%%%%%%%%%%%%%%%

\subsection{Axion examples\label{axionsect}}

Axion models may have an effective Chern-Simons coupling
\be
\lambda C_{ij} \cF_{tj} B
\ee
which naively generates damping at large $\lambda$.  
The effective action takes the form
\bea
\label{axionAct}
& &\cS_{\eff} = \int\! dt \biggl(\frac{a_1a_2a_3}{N} \Bigl[-\mpl^2\Bigl(\frac{\dot a_2\dot a_3}{a_2a_3}+{\it cyclic}\Bigr) +\frac{\gamma_c}{2} \frac{\partial_t(a_1a_2 c)^2}{a_1^2a_2^2}
	 + \frac{\gamma_\psi}{2 } \frac{\partial_t(a_3 \psi)^2}{a_3^2}
	 +  \frac{\gamma_b}{2}{\dot b}^2\Bigr]  \nonumber\\
		& & \hskip 4cm -Na_1a_2a_3 \cV(b) - \lambda a_1a_2cb\partial_t (a_3\psi)\biggr) \ ,
\eea
where $C_{12}=a_1a_2c$, $\cA_3=a_3\psi$, and the axion is $B_{\alpha\beta}=b$;
in the terminology of section\pref{CSgeneralities} one has a Chern-Simons coupling with $m=1$.  Effectively, in the Chern-Simons coupling of chromo-natural inflation, the commutator of gauge fields $[\cA_i,\cA_j]$ has been replaced by the antisymmetric tensor $C_{ij}$, and the gauge field made abelian.  Can this Chern-Simons coupling prolong inflation?

The equations of motion in the slow-roll approximation \pref{slowrollkinetics} become
\bea
\label{CBAeoms}
\gc (H_{tot}\dot c + H_3(H_1+H_2) c) &=& \lambda b( \dot \psi+H_3\psi) \nonumber\\
\gp (H_{tot}\dot \psi + H_3(H_1+H_2) \psi) &=& -\lambda( c\dot b + b\dot c +(H_1+H_2)bc)\\
\gb (H_{tot}\dot b) +\cV' &=&  \lambda c( \dot\psi+H_3\psi) \nonumber
\eea
with $H_{tot}=H_1+H_2+H_3$.  Subtracting $c$ times the first equation from $b$ times the third,
\be
\label{bcdiff}
H_{tot}(\gc c\dot c-\gb b\dot b)+H_3(H_1+H_2)\gc c-b\cV'(b) = 0 \ ,
\ee
we find that generically $c$ is of order $\lambda^0$.  Assuming that the velocities are damped by inverse powers of $\lambda$ fixes $c$ in terms of $b$.  Proceeding to diagonalize the velocities as before, one finds that a quasi-static solution for the $\psi$ dynamics fixes
\be
\psi \approx \frac{\gc c\bigl(\cV'(b)-\gb b(H_1+H_2)^2\bigr)}{\lambda H_3(\gb b^2+\gc c^2)} \ ,
\ee
so that $\psi\sim \lambda^{-1}$.  Putting this back into~\pref{bcdiff}, one finds that if the velocities are $\lambda$-suppressed, either $b=0$, or $c=0$, or $H_3=-(H_1+H_2)$.  Either way, the dynamics is not governed by a self-consistent inflationary solution where velocities are controlled by an expansion in inverse powers of $\lambda$. 
One is led to conclude that only $m=0$ or $m=3$ in the general Chern-Simons term~\pref{CSstruct} gives rise to inflationary dynamics.

%%%%%%%%%%%%%%%%%%%%%%%%%%%%%%%%%%%%%%%%%%%%%%%
%%%%%%%%%%%%%%%%%%%%%%%%%%%%%%%%%%%%%%%%%%%%%%%

\section{A preliminary look at fluctuations\label{perts}}

One of the great successes of inflation is the prediction of a nearly scale-invariant spectrum of density fluctuations (for a review, see for example%
~\cite{Mukhanov:1990me})
in beautiful agreement with observation%
~\cite{Komatsu:2010fb}.  
In this section, we present some preliminary results on the fluctuation spectrum of EM-flation, in the D7-brane example of section~\ref{D7sect}.

We will ignore the panoply of tensor fluctuations and focus on the fluctuations of the three actors in the Chern-Simons term.  Their equations of motion are
\bea
\Bigl[\partial_t^2+H_{tot}\partial_t +\frac{k_i^2}{a_i^2}+H_3(H_1+H_2)+\dot H_3\Bigr]\delta \dc &=& \bflam[\dot \dX\delta\db + \bhs\partial_t \delta \dX] \\
\Bigl[\partial_t^2+H_{tot}\partial_t +\frac{k_i^2}{a_i^2}+H_3(H_1+H_2)+(\dot H_1+\dot H_2)\Bigr]\delta \db &=& \bflam[\dot \dX\delta\dc + \chs\partial_t \delta \dX] \\
\Bigl[\partial_t^2+H_{tot}\partial_t +\frac{k_i^2}{a_i^2}+\cV''(\dX)\Bigr]\delta \dX &=& -  \bflam[(\partial_t+H_{tot}) (\chs\delta \db+\bhs\delta\dc)]\hskip 1cm
\eea
with $\bflam =\lambda/\sqrt{\gamma_b\gamma_c\gamma_X}$,  and the canonically normalized fields $\dc = \sqrt{\gamma_c} \, c$, $\db=\sqrt{\gamma_b} \,b$, $\dX=\sqrt{\gamma_X}\,X$; also $H_{tot}=H_1+H_2+H_3$.  Taking the difference of the first two equations, dropping the $\dot H_i$ terms, and using the fact that $\bhs=\chs$ at leading order in $1/\lambda$, we see that $(\db-\dc)$ is exponentially damped to zero; we will henceforth ignore it.  The other two equations are less trivial:
\bea
\Bigl[\partial_t^2+H_{tot}\partial_t +\frac{k_i^2}{a_i^2}+H_3(H_1+H_2)\Bigr]\delta( \db+\dc)&=& \bflam[\dot \dX\delta( \db+\dc) + 2\bhs\partial_t \delta \dX] \\
\Bigl[\partial_t^2+H_{tot}\partial_t +\frac{k_i^2}{a_i^2}+\cV''(\dX)\Bigr]\delta \dX &=& - \bflam{\bhs}[(\partial_t+H_{tot} )\delta( \db+\dc)]
\eea
At this point we will make some simplifications.  First note that, up to order $1/\lambda$, all the $H_i$ are equal to $H_{tot}/3$, and on the classical solution~\pref{Xvel} to leading order in $\lambda$ we have $\bflam\dot \dX\approx 2H^2$, so that the nonderivative terms proportional to $\delta(\db+\dc)$ cancel in the first equation (\ie\ the residue is of order $1/\lambda$).  Eventually we will want to treat the full system, get the tilt and anisotropy of the power spectrum, analyze tensor perturbations, and so on.  However, for a first attempt, we will simply freeze the background to isotropic de Sitter space with some given Hubble constant $H$, and analyze these fluctuation equations to look for a scale-invariant spectrum.  

The structure of these equations is encouraging -- we have two light fields, and the job of the RHS is to trade energy between the two.  However, one might worry that the inflaton potential's curvature $\cV''$ is not small, and so might lead to anomalously large fluctuations or even instability.  Instead, we will see that~-- as in the case of the evolution of the background~-- the effect of the Chern-Simons terms on the RHS tames the fluctuation spectrum as well, leading to a result quite analogous to the single-field model.

Let us pass to conformal time
\be
a\,d\tau =  dt~~,\quad a=-1/(H\tau)~~,\quad \partial_t = -H\tau\partial_\tau
\ee
and define $\beta=a(\tau)\delta( \db+\dc)$ and $\chi=a(\tau)\delta \dX$.
To leading order in $1/\lambda$, one finds
\bea
\partial_\tau^2\beta+k^2 \beta-\frac{2}{\tau^2} \beta &=& - \frac{\bflam\db^*}{H}\Bigl[\frac{2\partial_\tau\chi}{\tau} +\frac{2\chi}{\tau^2}\Bigr] \nonumber\\
\partial_\tau^2\chi+k^2\chi-\frac{1}{\tau^2}\Bigl(2-\frac{\cV''}{H^2}\Bigr)\chi &=& \frac{\bflam\db^*}{H}\Bigl[\frac{\partial_\tau\beta}{\tau} -\frac{2\beta}{\tau^2}\Bigr] \ .
\eea
It is not yet clear whether this system admits analytical solution, but the asymptotics are rather straightforward.   For late time $k\tau\gtrsim -1$, the equations are solved by power laws
\be
\chi\sim C_\chi \tau^{\alpha}~~,\quad \beta\sim C_\beta \tau^{\alpha}
\ee
so, defining $L=\bflam\bhs/H$, and $\nu=\cV''/H^2$, we have
\bea
\bigl( \alpha(\alpha-1)-2 \bigr)C_\beta &=& L(-2\alpha-2)C_\chi \nonumber\\ 
\bigl( \alpha(\alpha-1)-2+\nu \bigr) C_\chi &=& L(\alpha-2)C_\beta \ .
\eea
There are four roots for $\alpha$:
\be
\label{alpharoots}
 -1~,~~ 2 ~,~~
\frac{1}{2} \left(1+i\sqrt{8 L^2+4 \nu -9}\right) ~,~~
\frac{1}{2} \left(1-i\sqrt{8 L^2+4 \nu-9}\right)~~.
\ee
There is always a growing mode with $\alpha=-1$, independent of the curvature of the potential $\nu$ at this order.  This mode has $C_\chi=-(3  L/\nu)C_\beta$, so the bulk of the power is in the inflaton $\dX$ with only an amount of order $1/\lambda$ in the perturbations of the AST fields $\db$, $\dc$.  The remaining modes are damped out at late time and consequently irrelevant.

For early times $k\tau\ll -2L$, we have a free wave equation for both $\chi$ and $\beta$.   In fact we can do better; for $k\tau\ll-1$, we can ignore all the $1/\tau^2$ terms as being down by powers of $k\tau$, resulting in
\bea
\partial_\tau^2\beta + k^2\beta &=& -\frac{2L}{\tau} \partial_\tau \chi \nonumber\\
\partial_\tau^2\chi + k^2\chi &=& +\frac{2L}{\tau} \partial_\tau \beta 
\eea
which is our friend from section~\ref{toy}, the charged particle in a (now time-dependent) magnetic field of strength $2L/\tau$, and harmonic well of frequency $k$.  The eigenmodes as before are circular motions with angular momentum in the same/opposite orientation as the magnetic field; the derivative terms are diagonalized by the linear combinations $\Psi_\pm=\chi\pm i\beta$, each of which has two modes -- a fast mode and a slow mode
\be
\label{toymodes}
\Psi_\pm = \chi\pm i\beta = 
C_\rs^\pm\exp[+i\phi_\rs^\pm(\tau)] + C_\rf^\pm \exp[+i\phi_\rf^\pm(\tau)] \ ,
\ee
so-called because in the case of large {\it constant} magnetic field the fast mode oscillates at the Larmor frequency and the slow mode undergoes magnetic drift.
The WKB eikonals are
\bea
\phi^\pm_\rf &=& \pm\Bigl( +\sqrt{k^2 \tau^2+L^2}-L \log \bigl(\sqrt{k^2 \tau ^2+L^2}+L\bigr)+2 L \log (\tau) \Bigr) \nonumber \\
\phi^\pm_\rs &=& \pm\Bigl( -\sqrt{k^2 \tau ^2+L^2}+L \log \bigl(\sqrt{k^2 \tau ^2+L^2}+L\bigr) \Bigr) \ .
\eea

It seems clear that the fast modes, whose small $\tau$ behavior has an extra phase oscillation $\tau^{\pm i2L}$, are designed to match onto the late time phase oscillation $\tau^{\pm i \sqrt2 L}$ of the last, conjugate pair of roots of $\alpha$ in~\pref{alpharoots}.  The real roots for $\alpha$ in the late time solution then match onto the slow modes of the early time solution.

The conventional analysis of single-field inflation sets the Bunch-Davies vacuum as an initial state, then lets the wavefunction evolve.  The late time solution is a constant of order one times $H$ (taking into account the factor of $a$ that was put into the perturbations upon going to conformal time); the detailed form of the evolution determines that constant of order one, thus providing the absolute normalization of the power spectrum.  

The Bunch-Davies vacuum consists of setting to zero the amplitudes of the modes behaving as $e^{+ik\tau}$ at early times, \ie\  $C^+_\rf=C^-_\rs=0$; and thus $C^-_\rf=C^+_\rs=1$.  The matching of the early and late time asymptotics will determine the normalization of the fluctuations in the present situation as well, and therefore that of the power spectrum.  While we have not yet determined the constant of order one by the matching procedure, it is clear that the normalization is again a factor of $H$ times such a constant.  Note that the curvature of the potential $\nu$ has dropped out of the asymptotics of the slow mode, the one that leads to growing perturbations on scales beyond the de Sitter horizon.

It is remarkable in the present context that the Chern-Simons interaction leads to a fluctuation amplitude that appears to be largely independent of the details of the shape of the potential, only depending on its overall scale.  In this model there are three scales, the scale $\mu$ of the potential, the scale $R$ of the compactification, and the string scale $\lstr$. From these we have two dimensionless ratios, as well as the string coupling $\gstr$ and number of branes $N$.  We can use these quantities to set the the duration of inflation to be $O(60)$ e-foldings, and the scale of the fluctuations to be $O(10^{-5})$.  Feeding the normalization $H$ of the fluctuation amplitude into the curvature perturbation amplitude $\cR$ (in a gauge with spatially flat slicing),
\be
\cR \approx \frac{\dot\dX \delta\dX+H(\dc\delta\dc+2\db\delta\db)}{2\mpl^2 H\epsilon_H} \ ,
\ee
we find for this D7-brane example
\be
\label{curlyR}
\cR \sim N^{1/2}\gstr^{5/2} \Bigl(\frac{\lstr}{R}\Bigr)^{10}\,\mu^2 R^2\;  \frac{f^{3/2}}{f'} \ .
\ee
When rewritten in four-dimensional terms, this expression is the ratio of two small quantities and therefore is indeterminate.  Expressed in ten-dimensional terms, the fluctuation spectrum gives us direct constraints on the dimensionless parameters of the theory.
The observed value is $\cR\sim 10^{-5}$; recalling that the number of e-foldings~\pref{efolds} is $N_e\sim N^{1/2}/(\mu^2R^2)$, we see that 
$ N^{3/2}\gstr^{5/2}({\lstr}/{R})^{10}$ has to be of the order $10^{-3}$.  This is for instance satisfied for $N=1$, $R\sim \lstr$ and $\gstr\sim 10^{-1}$, with $\mu \sim 10^{-1}R^{-1} \sim 10^{-2}\mpl$.

This particular example thus seems to favor high scale inflation.  If all the dimensions of the compactification are roughly similar, the curvature perturbation amplitude~\pref{curlyR} is such a steep function of $R$ that it is difficult to make the size of the compactification much more than the string scale, which is quite close to the ten-dimensional Planck scale.  This result might favor a warped compactification scenario%
~\cite{Kachru:2003sx}
where one dimension is somewhat larger than the others, in order to give the branes some room to roam.

%%%%%%%%%%%%%%%%%%%%%%%%%%%%%%%%%%%%%%%%%%%%%%%
%%%%%%%%%%%%%%%%%%%%%%%%%%%%%%%%%%%%%%%%%%%%%%%

\section{Discussion}

We have outlined a new mechanism of inflation based on the idea that a Chern-Simons term has the same effect on the effective field space dynamics as a magnetic field does on a charged particle; inflation can then be slow-roll because the inflaton experiences `magnetic drift' forces that balance the force from the inflaton potential, when the effective Chern-Simons coupling is parametrically large.  In such a situation, the kinetic term of the effective inflaton plays only a minor role, as the forcing from the potential is compensated largely by the Chern-Simons term.  Thus there should be no eta problem; it is not the kinetic term that matters, nor the flatness of the potential, but rather the strength of the Chern-Simons term, which is quasi-topological in nature and therefore rather insensitive to quantum corrections in the effective action.  
The Chern-Simons coupling is of order one in string units (or Planck units, in M-theory), times an integer; so long as the scale of the inflaton potential is sufficiently small relative to this scale and the compactification scale, a long period of slow-roll inflation results.  The examples investigated above bear out this intuition.

A general feature of this inflationary scenario is its robustness against quantum corrections to the effective action of the inflaton, which is the source of the eta problem in other approaches.  Quantum effects generate corrections to the effective potential and kinetic terms.  These corrections spoil any delicate tuning that has been done in order to engineer a long period of inflation, when only these terms in the action are active in the inflationary mechanism.  In EM-flation, it doesn't matter whether there are order one corrections of this sort -- the expressions~\pref{etaeps} for the slow roll parameters are indeed order one ratios of the effective potential and its derivatives, times the velocity of the inflaton.  However, this velocity is of order $1/\lambda$ due to the influence of `magnetic damping', instead of the value that pertains when only potential forces and Hubble damping are present.  By a suitable arrangement of the scales in the problem, and perhaps tuning some discrete parameters, the slow-roll parameters are reliably kept small.

How large is the EM-flation basin of attraction?  Consider the D7-brane example of section~\ref{D7sect}.  
In the large $\lambda$ limit, the evolution of $b$, $c$ is much faster than $X$, so we take the latter as frozen at some value during the time that $b$ and $c$ are relaxing to an attractor.  The equations of motion for these variables have the form
\bea
\Gamma(w)\partial_t w_a &=& \partial_{w_a} \cU(w) \nonumber\\
\cU &=&(\hat\lambda\cV') w_1 w_2 + \frac16 \hat\lambda^2 H(w_1^4+w_2^4+7 w_1^2w_2^2)+3\gamma_X H^3(w_1^2+w_2^2) \nonumber\\
\Gamma &=& 9\gamma_X H^2 + \hat\lambda^2(w_1^2+w_2^2) 
\eea
where $w_1=\sqrt\gamma_b\, b$, $w_2=\sqrt{\gamma_c}\,c$, and $\hat \lambda = \lambda/\sqrt{\gamma_a\gamma_b}$.  The slow-roll dynamics of $b$, $c$ at fixed $X$ is gradient flow, and a plot of the potential $\cU$ shows that the fixed point at $b=c=0$ is an unstable saddle point, while EM-flation is the only stable attractor of the dynamics.  The basic intuition is that, once all the fields are nonzero, the Chern-Simons term is active in the dynamics and the dynamics of the inflaton $X$ drive the AST fields $b$, $c$ to larger values and the EM-flation attractor.  So in any example where the EM-flation mechanism is in effect, it applies to {\it all} inflationary trajectories.  Similarly, the attractor for chromo-natural inflation is the dominant component of the gauge field configuration space, with only a small region around the origin (whose size scales inversely with $\lambda$) attracted toward $\cA=0$ rather than the chromo-natural attractor.

Finally, it seems that a characteristic feature of the EM-flation model of section~\ref{D7sect} is a slight anisotropy of the Hubble expansion rate during inflation, see for example equation~\pref{deltaH}, driven by an expectation value for AST fields.  This anisotropy will imprint on the CMB perturbations and lead to a potentially observable signature.   In the context of ordinary inflation, the effect of an anisotropic Hubble expansion rate was considered in%
~\cite{Ackerman:2007nb}, 
with the result that the power spectrum picks up a rotationally non-invariant contribution $\delta\cP/\cP$ of order $(H_3-H_2)/H_{tot}$.  This latter quantity was estimated in the example of section~\pref{D7sect} to be generically of order $\epsilon_H/\sqrt{N}$.  The fluctuation spectrum anisotropy $\delta\cP/\cP$ might be measured down to the level of one percent by the Planck experiment%
~\cite{Ma:2011ii}.%
\footnote{An initial indication of such an effect in the data%
~\cite{Bennett:2010jb}
has since been discounted as an instrumental effect%
~\cite{Hanson:2010gu}.}
It seems reasonable that the fluctuation spectrum anisotropy of EM-flation is at a level similar to that found in%
~\cite{Ackerman:2007nb} 
for ordinary inflation, given the similarity of the perturbation calculation of section~\ref{perts}; then a fluctuation anisotropy signal might well be within experimental reach.  Needless to say, the detection of such an anisotropy would constitute strong evidence for string theory.

\vskip 2cm
\noindent{{\bf Acknowledgments:}}
Thanks to J. Harvey, W. Hu and J. Marsano for helpful discussions, and to E. Silverstein for comments on the first version of this article.  This work was supported in part by DOE grant DE-FG02-90ER-40560, and by the Kavli Institute for Cosmological Physics at the University of Chicago through NSF grants NSF PHY-0114422, NSF PHY-0551142 and an endowment from the Kavli Foundation and its founder Fred Kavli.
P.A. and M.W. thank the Aspen Center for Physics for its hospitality and support through National Science Foundation Grant No. 1066293, as this work was nearing completion.

%%%%%%%%%%%%%%%%%%%%%%%%%%%%%%%%%%%%%%%%%%%%%%%
%%%%%%%%%%%%%%%%%%%%%%%%%%%%%%%%%%%%%%%%%%%%%%%

\vskip 2cm

%\newpage
\bibliographystyle{amsunsrt-ensp}
\bibliography{EMflation}
\end{document}